\documentclass[aps,prd,amsmath,two column,amssymb,showpacs]{revtex4}
\usepackage{amssymb}
\usepackage{txfonts}
\usepackage{mathbbold}
\usepackage{amsfonts}
\usepackage{mathrsfs}
\usepackage{epsfig,bm,dcolumn}
\usepackage{graphicx}
\usepackage{color}
\usepackage{amsmath}
\usepackage{dcolumn}
\usepackage{overpic}

\newcommand{\feyn}[1]{\setbox0=\hbox{\ensuremath{#1}}\hbox to\wd0{\hbox to0pt{\hbox to\wd0{\hss/\hss}\hss}\box0}}

\begin{document}

\title{Collective modes and Kosterlitz-Thouless transition in a magnetic field \\ in the planar Nambu--Jona-Lasino model}

\author{Gaoqing Cao,$^{1}$ Lianyi He,$^{2,3}$ and Pengfei Zhuang$^{1}$}
\affiliation{1 Department of Physics, Tsinghua University and Collaborative Innovation Center of Quantum Matter, Beijing 100084, China\\
2 Theoretical Division, Los Alamos National Laboratory, Los Alamos, NM 87545, USA\\
3 Frankfurt Institute for Advanced Studies, 60438 Frankfurt am Main, Germany}

\date{\today}

\begin{abstract}
It is known that a constant magnetic field is a strong catalyst of dynamical chiral symmetry breaking in 2+1 dimensions, leading to generating dynamical fermion mass even at weakest attraction. In this work we investigate the collective modes associated with the dynamical chiral symmetry breaking in a constant magnetic field in the (2+1)-dimensional Nambu--Jona-Lasinio model with continuous U(1) chiral symmetry. We introduce a self-consistent scheme to evaluate the propagators of the collective modes at the leading order in $1/N$. The contributions from the vacuum and from the magnetic field are separated such that we can employ the well-established regularization scheme for the case of vanishing magnetic field. The same scheme can be applied to the study of the next-to-leading order correction in $1/N$. We show that the sigma mode is always a lightly bound state with its mass being twice the dynamical fermion mass for arbitrary strength of the magnetic field. Since the dynamics of the collective modes is always 2+1 dimensional, the finite temperature transition should be of the Kosterlitz-Thouless (KT) type. We determine the KT transition temperature $T_{\rm KT}$ as well as the mass melting temperature $T^*$ as a function of the magnetic field. It is found that the pseudogap domain $T_{\rm KT}<T<T^*$ is enlarged with increasing strength of the magnetic field. The influence of a chiral imbalance or axial chemical potential $\mu_5$ is also studied. We find that even a constant axial chemical potential $\mu_5$ can lead to inverse magnetic catalysis of the KT transition temperature in 2+1 dimensions. The inverse magnetic catalysis behavior is actually the de Haas--van Alphen oscillation induced by the interplay between the magnetic field and the Fermi surface.
\end{abstract}


\maketitle

\section {Introduction}

Dynamical chiral symmetry breaking plays a crucial role in understanding the ground state and particle spectroscopy of Quantum Chromodynamics (QCD) \cite{Peskin}. For example, the lightest mesons in the QCD spectra, the pions, are identified as pseudo-Goldstone bosons associated with the dynamical chiral symmetry breaking. Dynamical chiral symmetry breaking is also important for us to understand the phase structure of strongly interacting matter in extreme conditions, e.g., at high temperature and/or baryon density \cite{ReviewNJL1,ReviewNJL2,ReviewNJL3,ReviewNJL4,CSCReview1,CSCReview2,CSCReview3,PhaseReview}. It is generally believed that the broken chiral symmetry gets restored at high temperature and/or density. In general, dynamical chiral symmetry breaking is characterized by the nonzero expectation value $\langle\bar{\psi}\psi\rangle$, where $\psi$ denotes the quark field. The chiral symmetry breaking and its restoration at finite temperature/or density can be successfully described by some QCD motivated effective models, such as the Nambu--Jona-Lasinio (NJL) model \cite{NJL}.

Good knowledge of QCD in extreme conditions is therefore important for us to understand a wide range of physical phenomena \cite{PhaseReview}. For example, to understand the evolution of the early Universe in the first few seconds, the nature of the QCD phase transition at high temperature
and nearly vanishing baryon density is needed. On the other hand, to understand the physics of compact stars, we need the knowledge of the equation of state and dynamics of QCD matter at high baryon density and low temperature. In recent years, the phase structure of QCD matter in strong magnetic field $B$ promoted great interests \cite{BQCD-LSM,BQCD-SC,BQCD-NJL,BQCD-FRG,BQCD-CME,BQCD-CSC}. A strong magnetic field $B$ can be realized in non-central heavy ion collisions at the Relativistic Heavy-Ion Collider (RHIC) and the Large Hadron Collider (LHC). Some calculations have estimated that the produced magnetic field can be as large as $\sqrt{eB}\sim\Lambda_{\rm QCD}$ at the RHIC energy \cite{Bstrength}. At the LHC energy, even stronger $B$ can be produced.  On the other hand, the great theoretical advantage is that there is no sign problem for the Monte Carlo simulation of QCD at finite $B$. The lattice simulation of QCD at finite temperature and magnetic field $B$ has been performed with almost physical quark masses \cite{Lattice-B01,Lattice-B02}. It has been found that the transition temperature decreases with increasing magnetic field up to $\sqrt{eB}\simeq 1$GeV. Some theoretical explanations for this phenomenon (called inverse magnetic catalysis) have been proposed \cite{Inverse01,BInhibition,Inverse02,Inverse03,Inverse04,Inverse05,Inverse06}.

The effects of magnetic fields on the dynamical chiral symmetry breaking have been extensively studied in (2+1)- and (3+1)-dimensional four-fermion interaction models \cite{NJL-B01,NJL-B02}. In the absence of magnetic fields, dynamical chiral symmetry breaking occurs only when the four-fermion coupling strength is larger than a critical value, which is known as a quantum critical phenomenon \cite{NJL2DReview}. In the presence of a constant magnetic field, it was first shown by Klimenko and by Gusynin, Miransky, and Shovkovy that to the leading order of the large-$N$ expansion the magnetic field plays the role of a strong catalysis of dynamical chiral symmetry breaking, leading to generating a dynamical fermion mass even at the weakest attraction \cite{NJL-B01,NJL-B02}. For four-fermion coupling stronger than the critical value, the magnetic field enhances the dynamical chiral symmetry breaking and hence the dynamical fermion mass. This phenomenon is called magnetic catalysis \cite{NJL-B02}. To understand the underlying physics, we note that the low energy dynamics of pairing fermions undergoes dimension reduction $D\rightarrow D-2$ (at the lowest Landau level) in strong magnetic field, where $D$ is the space-time dimension of the system.

On the other hand, mesonic collective modes (the massive $\sigma$ mode and the Goldstone pion mode) should appear associated with the spontaneous breaking of the continuous chiral symmetry. The influence of a constant magnetic field on the low energy spectra of the collective modes at leading order in $1/N$ was studied by Gusynin, Miransky, and Shovkovy by using the method of low energy expansion \cite{NJL-B02}. The magnetic field strongly affects the low energy spectra of the collective modes even though these modes are electrically neutral. The dynamics of the collective modes is still 2+1 dimensional even at strong magnetic field, in contrast to the dynamics of the fermions. However, to our knowledge, so far a self-consistent scheme to study the full spectra of the collective modes is still missing. For example, the properties of the sigma mode obtained from the low energy expansion method cannot reveal the fact that the sigma mode is a lightly bound state with its mass equal to twice the dynamical fermion mass. This inconsistency can be attributed to the commonly used regularization scheme where a lower cutoff for the Schwinger parameter is introduced. Such a regularization scheme is proper to study the dynamical fermion mass and the low energy spectrum of the Goldstone mode. Inconsistency arises if we evaluate the full propagators of the collective modes at leading order in $1/N$. Different cutoffs should be used to make the Goldstone mode propagator compatible with the gap equation and therefore the Goldstone theorem \cite{ColMode}. Moreover, such a scheme becomes improper if we try to study the next-to-leading order corrections in $1/N$ \cite{NJL-NLO}.

In the first part of this paper, we employ a self-consistent scheme to evaluate the full propagators of the collective modes at leading order in $1/N$. Following the treatment of Klimenko \cite{NJL-B01}, we separate the leading-order effective potential into the vacuum contribution and the contribution from the magnetic field. Since the contribution from the magnetic field is finite, we can employ the usual regularization scheme which is used at vanishing magnetic field, where a cutoff for the Euclidean momentum is introduced. Note that this usual regularization scheme will be helpful if we need to calculate the next-to-leading order corrections. It is expected that the next-to-leading order corrections in $1/N$ (contributions from the collective modes) will be significant at strong magnetic field. It was shown in 3+1 dimensions that the next-to-leading order contributions generally lead to an opposite effect, called magnetic inhibition \cite{BInhibition}, which suppresses the magnetic catalysis effect. For a realistic system with small $N$, the inhibition effect may become competitive with or even dominant over the catalysis effect.

In the large-$N$ limit, phase fluctuations of the order parameter are completely suppressed and the system undergoes a second-order phase transition at a critical temperature where the dynamical fermion mass vanishes. However, for finite $N$, the CMWH theorem forbids any long-range order and hence spontaneous breaking of the U$(1)$ chiral symmetry at any nonzero temperature \cite{CMWH}. Since the dynamics of the collective modes is 2+1 dimensional, the finite temperature transition at finite $N$ should be of the Kosterlitz-Thouless (KT) type \cite{KT}. The KT transition temperature of the 2+1 dimensional  Nambu--Jona-Lasino model at vanishing magnetic field was studied by Babaev \cite{Babaev}. In the second part of this paper, we study the influence of a constant magnetic field on the KT transition temperature. The effect of the chiral imbalance will also be studied. We will show that even a constant axial chemical potential leads to inverse magnetic catalysis of the KT transition temperature in 2+1 dimensions. The inverse magnetic catalysis behavior can be attributed to a reflection of the de Haas--van Alphen oscillation \cite{dHvA}.

The paper is organized as follows. We set up the model and study the magnetic catalysis and collective modes at zero temperature in Sec. \ref{s2}.
The KT transition and influence of the chiral imbalance are investigated in Sec. \ref{s3}. We summarize in Sec. \ref{s4}.

\section {Zero temperature: magnetic catalysis and collective modes} \label{s2}
The Lagrangian density of the (2+1)-dimensional Nambu--Jona-Lasinio (NJL$_3$) model is given by \cite{NJL2DReview}
\begin{eqnarray}\label{GN2}
{\cal L}=\bar{\psi}i\feyn{\partial}\psi
+\frac{G}{2N}\left[(\bar{\psi}\psi)^{2}+(\bar{\psi}i\gamma_{5}\psi)^{2}\right],
\end{eqnarray}
where $\psi=(\psi_1,\psi_2,\ldots,\psi_N)$ denotes the $N$-flavor fermion fields with each $\psi_i$ being a four-component spinor and $G$ is the coupling constant. The $\gamma$-matrices are $4\times4$ matrices and can be defined as \cite{Gamma2D}
\begin{eqnarray}\label{gamma}
&&\gamma^0=\left(\begin{array}{cc} \sigma_3& 0\\ 0& -\sigma_3\end{array}\right),\ \ \
\gamma^1=\left(\begin{array}{cc} i\sigma_1& 0\\ 0& -i\sigma_1\end{array}\right),\nonumber\\
&&\gamma^2=\left(\begin{array}{cc} i\sigma_2& 0\\ 0& -i\sigma_2\end{array}\right),\ \ \
\gamma^5=i\left(\begin{array}{cc} 0& I\\ -I& 0\end{array}\right).
\end{eqnarray}
Here $\sigma_1,\sigma_2,$ and $\sigma_3$ are $2\times2$ Pauli matrices and $I$ is the $2\times2$ identity matrix. Note that the matrix $\gamma^5$ anticommutes with $\gamma^0,\gamma^1,$ and $\gamma^2$. The NJL$_3$ model is symmetric under the continuous chiral transformation $\psi_i\rightarrow e^{i\theta\gamma_5}\psi_i$. Spontaneous breaking of the chiral symmetry in this model therefore leads to massless bosonic excitation, i.e., the Goldstone mode.
We assume the fermions are electrically charged with a uniform charge $e$ and there is an external constant magnetic field ${\bf B}$ perpendicular to the planar system. To couple the fermions with the magnetic field, we replace the derivative $\partial_\mu$ by the covariant derivative $D_\mu=\partial_\mu-ieA_\mu$, where $A_0=0$ and ${\bf A}=(0,Bx_1)$. Without loss of generality, we set $eB>0$ in this paper.

\subsection {Effective potential and magnetic catalysis}

The calculation of the effective potential can be performed in the $1/N$ expansion. For the NJL$_3$ model, we introduce two auxiliary fields, $\sigma$ and $\pi$. The partition function reads
\begin{widetext}
\begin{eqnarray}
Z=\int[d\bar{\psi}][d\psi][d\sigma][d\pi]\exp\left\{i\int d^3x \left[\bar{\psi}(i\feyn{D}-\sigma-i\gamma_5\pi)\psi
-\frac{N}{2G}(\sigma^2+\pi^2)\right]\right\}.
\end{eqnarray}
Integrating out the fermion fields and introducing external sources $J_\sigma$ and $J_\pi$, we obtain the generating functional $W[J]$,
\begin{eqnarray}
&&Z[J]=e^{iW[J]}=\int[d\sigma][d\pi]\exp\left\{i\int d^3x \Big[{\cal L}_{\rm B}(\sigma,\pi)+J_\sigma\sigma+J_\pi\pi\Big]\right\},\nonumber\\
&&\int d^3x{\cal L}_{\rm B}(\sigma,\pi)=-iN{\rm Trln}(i\feyn{D}-\sigma-i\gamma_5\pi)-\frac{N}{2G}\int d^3x(\sigma^2+\pi^2).
\end{eqnarray}
\end{widetext}
The classical fields are given by
\begin{eqnarray}
\sigma_{\rm cl}=\left.\frac{\delta W[J]}{\delta J_\sigma}\right|_{J_\sigma,J_\pi=0},\ \ \ \ \pi_{\rm cl}=\left.\frac{\delta W[J]}{\delta J_\pi}\right|_{J_\sigma,J_\pi=0}.
\end{eqnarray}
Since the Lagrangian of the NJL$_3$ model is symmetric under the U$(1)\times$U$(1)$ chiral transformation, the effective potential should only depends on the combination $\sigma_{\rm cl}^2+\pi_{\rm cl}^2$. We can therefore choose $\sigma_{\rm cl}=M$ and $\pi_{\rm cl}=0$ without loss of generality. The quantity $M$ serves as the order parameter of spontaneous chiral symmetry breaking. Then making the field shifts $\sigma=M+\tilde{\sigma}$ and $\pi=0+\tilde{\pi}$, we find that the $1/N$ expansion corresponds to the expansion in powers of the fluctuation fields $\tilde{\sigma}$ and $\tilde{\pi}$. To the next-to-leading order, the effective action $\Gamma(M)$ reads
\begin{eqnarray}
\Gamma(M)=N\Gamma^{(0)}(M)+\Gamma^{(1)}(M)+O\left(\frac{1}{N}\right),
\end{eqnarray}
where
\begin{eqnarray}
&&\Gamma^{(0)}(M)=-i{\rm Trln}(i\feyn{D}-M)-\int d^3x\frac{M^2}{2G},\nonumber\\
&&\Gamma^{(1)}(M)=\frac{i}{2}\ln\det\left[\frac{\delta^2{\cal L}_{\rm B}}{\delta\tilde{\sigma}\delta\tilde{\sigma}}\right]_{\tilde{\sigma},\tilde{\pi}=0}
+\frac{i}{2}\ln\det\left[\frac{\delta^2{\cal L}_{\rm B}}{\delta\tilde{\pi}\delta\tilde{\pi}}\right]_{\tilde{\sigma},\tilde{\pi}=0}.
\end{eqnarray}

The effective potential $\Omega(M)$ is given by $\Omega(M)=-\Gamma(M)/V_{2+1}$, where $V_{2+1}$ is the space-time
volume in (2+1) dimensions. To the next-to-leading order in the $1/N$ expansion, the effective potential $\Omega(M)$ can
be formally expressed
\begin{eqnarray}
\Omega(M)=N\Omega^{(0)}(M)+\Omega^{(1)}(M)+O\left(\frac{1}{N}\right).
\end{eqnarray}
The leading-order contribution in $1/N$ expansion is given by
\begin{eqnarray}
\Omega^{(0)}(M)=\Omega^{(0)}_0(M)+\Omega^{(0)}_B(M),
\end{eqnarray}
where the $B$-independent vacuum part $\Omega^{(0)}_0(M)$ reads
\begin{eqnarray}
\Omega^{(0)}_0(M)=\frac{M^2}{2G}-\frac{1}{\pi^2}\int_0^\Lambda dkk^2\ln(k^2+M^2).
\end{eqnarray}
Here and in the following we work in the Euclidean space for convenience. The vacuum part is divergent and we have introduced a cutoff $\Lambda$ for the Euclidean momentum $k$ to regularize the divergence. Neglecting the terms that are independent of $M$ and that vanish for $\Lambda\rightarrow\infty$, we obtain \cite{NJL2DReview}
\begin{eqnarray}
\Omega^{(0)}_0(M)=\frac{1}{2}\left(\frac{1}{G}-\frac{2\Lambda}{\pi^2}\right)M^2+\frac{M^3}{3\pi}.
\end{eqnarray}
The $B$-dependent part $\Omega^{(0)}_B(M)$ can be formally expressed as
\begin{eqnarray}
\Omega^{(0)}_B(M)=-\frac{1}{V_{2+1}}{\rm Trln}\frac{\feyn{D}+M}{\feyn{\partial}+M}.
\end{eqnarray}
This contribution is finite and we evaluate it by using the Schwinger approach \cite{Schwinger}. We get \cite{NJL-B01}
\begin{eqnarray}
\Omega^{(0)}_B(M)&=&\frac{1}{4\pi^{3/2}}\int_0^\infty\frac{ds}{s^{5/2}}e^{-sM^2}\left(\frac{eBs}{\tanh eBs}-1\right)\nonumber\\
&=&\frac{E_{\rm B}^3}{4\pi^{3/2}}f_{5/2}\left(\eta\right),
\end{eqnarray}
where $E_{\rm B}\equiv \sqrt{eB}$ is the energy associated with the magnetic field and $\eta\equiv M/E_{\rm B}$. The function $f_n(\eta)$ is defined as
\begin{eqnarray}
f_n(\eta)=\int_0^\infty\frac{dx}{x^n}e^{-\eta^2x}\left(\frac{x}{\tanh x}-1\right).
\end{eqnarray}

\begin{figure}[!htb]
\begin{flushright}
\includegraphics[width=8.3cm]{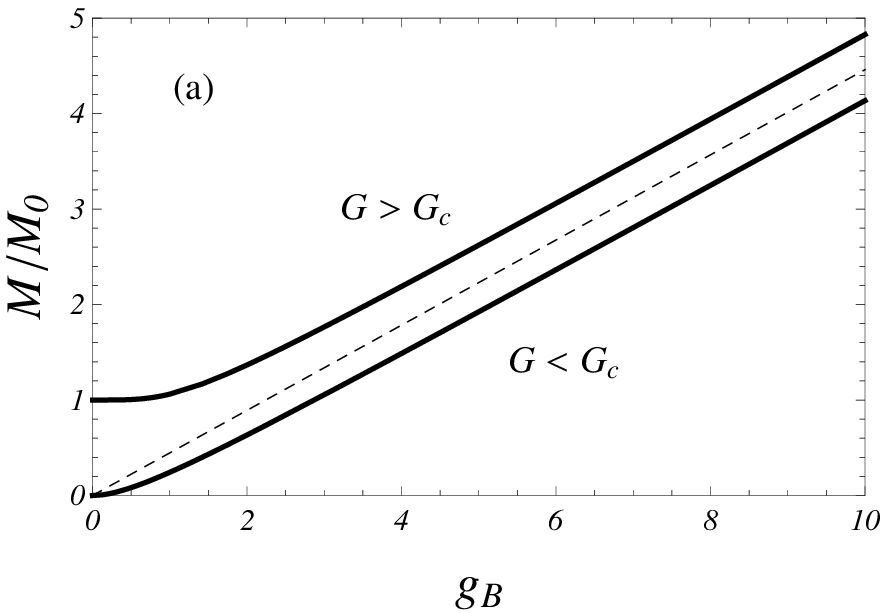}
\end{flushright}
\begin{center}
\includegraphics[width=8.5cm]{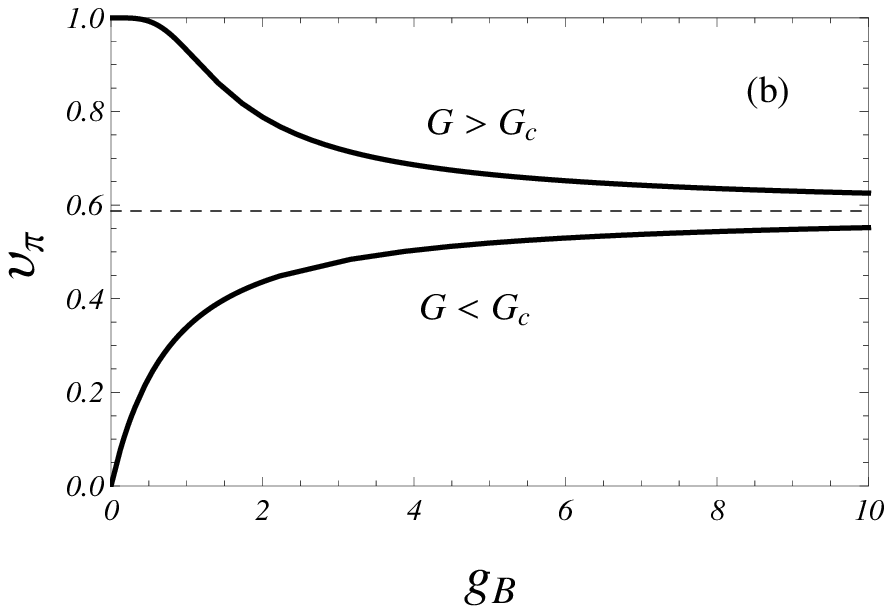}
\caption{(a) The dynamical fermion mass $M$ scaled by $M_0$ as a function of $g_{\rm B}=E_{\rm B}/M_0$ for the
supercritical case $G>G_c$ and subcritical case $G<G_c$. The dashed line denote the universal limit
$M/E_{\rm B}=0.4460$ for $B\rightarrow\infty$.
(b) The velocity of Goldstone mode $\upsilon_\pi$ as a function of $g_{\rm B}$ for the
supercritical case $G>G_c$ and subcritical case $G<G_c$. The dashed line denotes the universal limit
$\upsilon_\pi=0.5875$ for $B\rightarrow\infty$. \label{fig1}}
\end{center}
\end{figure}

The renormalization of the effective potential at leading order is simple. The bare coupling constant $G(\Lambda)$ should be fine-tuned such that \cite{NJL2DReview}
\begin{eqnarray}
\frac{1}{G(\Lambda)}-\frac{1}{G_c}=-\frac{M_0}{\pi}{\rm sgn}(G-G_c),
\end{eqnarray}
where the critical coupling $G_c=\pi^2/(2\Lambda)$ and $M_0>0$ is a finite quantity. The quantity $M_0$ then serves as a natural mass scale of the system. At $B=0$, spontaneous chiral symmetry breaking with $M\neq0$ is only possible when the coupling constant $G$ is larger than
the critical value $G_c$. The dynamical fermion mass reads $M=M_0$. However, in the presence of magnetic field, spontaneous chiral symmetry breaking occurs for arbitrarily weak coupling $G$. This can be seen from the fact that at $B\neq0$, $M=0$ is no longer a minimum of the effective potential $\Omega^{(0)}(M)$. Using the fact that
\begin{eqnarray}
\frac{1}{\sqrt{\pi}}\lim_{\eta\rightarrow0}\left[\eta f_{3/2}(\eta)\right]=1,
\end{eqnarray}
we obtain \cite{NJL-B02}
\begin{eqnarray}
\frac{d\Omega^{(0)}(M)}{dM}\Bigg|_{M=0}=-\frac{eB}{2\pi}.
\end{eqnarray}
Therefore, at the leading order of the $1/N$ expansion, we have the famous magnetic catalysis effect.

The gap equation that determines the dynamical fermion mass $M$ as a function of $E_{\rm B}$ can be expressed as
\begin{eqnarray}
\eta-\frac{1}{g_{\rm B}}{\rm sgn}(G-G_c)=\frac{1}{2\sqrt{\pi}}f_{3/2}\left(\eta\right).
\end{eqnarray}
Here the dimensionless parameter $g_{\rm B}\equiv E_{\rm B}/M_0$ which represents the strength of the magnetic field.The numerical results for the cases $G<G_c$ and $G>G_c$ are shown in Fig. \ref{fig1}. We find that $M$ is always an increasing
function of the magnetic field in both cases. In the strong magnetic field limit, the behavior of the dynamical fermion mass is universal.
For $g_{\rm B}\rightarrow\infty$, the universal ratio $\eta=M/E_{\rm B}$ is determined by the following equation
\begin{eqnarray}
\eta=\frac{1}{2\sqrt{\pi}}f_{3/2}\left(\eta\right).
\end{eqnarray}
We obtain in the strong magnetic field limit
\begin{eqnarray}
\lim_{B\rightarrow\infty}\frac{M}{E_{\rm B}}=0.4460.
\end{eqnarray}

\subsection {Collective modes}

At leading order in $1/N$, the propagators of $\sigma$ meson and pion read
\begin{eqnarray}
{\cal D}_\sigma(K)&=&\frac{1}{N}\frac{G}{1+G\Pi_\sigma(K)},\nonumber\\
{\cal D}_\pi(K)&=&\frac{1}{N}\frac{G}{1+G\Pi_\pi(K)},
\end{eqnarray}
where the polarization functions $\Pi_{\sigma,\pi}(K)$ are given by
\begin{eqnarray}
\Pi_\sigma(K)&=&\int\frac{d^3P}{(2\pi)^3}{\rm tr}\left[{\cal S}(P){\cal S}(P+K)\right],\nonumber\\
\Pi_\pi(K)&=&\int\frac{d^3P}{(2\pi)^3}{\rm tr}\left[{\cal S}(P)i\gamma_5{\cal S}(P+K)i\gamma_5\right].
\end{eqnarray}
Here ${\cal S}(P)$ is the fermion propagator up to a phase factor and is given by \cite{NJL-B02}
\begin{widetext}
\begin{eqnarray}
{\cal S}(P)=\int_0^\infty ds \exp\left[-s(M^2+p_3^2)-{\bf p}^2\frac{f_s}{eB}\right]
\left[M-\gamma_\mu P_\mu-i(p_2\gamma_1-p_1\gamma_2)f_s\right]\left(1-i\gamma_1\gamma_2f_s\right).
\end{eqnarray}
Here and in the following we use the notation $f_s=\tanh(eBs)$ for convenience.

To evaluate the  propagators of collective modes, we first complete the trace in the spin space and get
\begin{eqnarray}
\Pi_{\rm m}(K)=-\frac{4}{(2\pi)^3}\int_0^\infty ds\int_0^\infty dt e^{-(s+t)M^2-{\cal R}(s,t)}\int_{-\infty}^\infty dp_1 \int_{-\infty}^\infty dp_2 \int_{-\infty}^\infty dp_3e^{-{\cal A}(p_1,p_2,p_3)}{\cal G}_{\rm m}(p_1,p_2,p_3)
\end{eqnarray}
for ${\rm m}=\sigma,\pi$, where
\begin{eqnarray}
{\cal R}(s,t)=\frac{f_sf_t}{f_s+f_t}\frac{k_1^2+k_2^2}{eB}+\frac{st}{s+t}k_3^2,
\end{eqnarray}
\begin{eqnarray}
{\cal A}(p_1,p_2,p_3)=\frac{f_s+f_t}{eB}\left(p_1+\frac{f_t}{f_s+f_t}k_1\right)^2
+\frac{f_s+f_t}{eB}\left(p_2+\frac{f_t}{f_s+f_t}k_2\right)^2+(s+t)\left(p_3+\frac{t}{s+t}k_3\right)^2,
\end{eqnarray}
and
\begin{eqnarray}
{\cal G}_{\rm m}(p_1,p_2,p_3)=\left[\alpha_{\rm m}M^2+p_3(p_3+k_3)\right](1+f_sf_t)+\left[p_1(p_1+k_1)+p_2(p_2+k_2)\right](1-f_s^2)(1-f_t^2).
\end{eqnarray}
Here $\alpha_\sigma=-1$, $\alpha_\pi=1$. However, the integral over $P$ is divergent and we cannot simply shift the integration variables. To this end, we consider the combined quantity
$1/G+\Pi_{\rm m}(K)$, which is finite and hence independent of the cutoff $\Lambda$. We therefore use the following trick:
\begin{eqnarray}
\frac{1}{G}+\Pi_{\rm m}(K)=F_1+F_2+F_{\rm m}(K),
\end{eqnarray}
where
\begin{eqnarray}
&&F_1=\frac{1}{G}-4\int\frac{d^3P}{(2\pi)^3}\int_0^\infty ds e^{-s(M^2+P^2)}=\frac{1}{G}-4\int\frac{d^3P}{(2\pi)^3}\frac{1}{M^2+P^2},\nonumber\\
&&F_2=4\int\frac{d^3P}{(2\pi)^3}\int_0^\infty ds\left[e^{-s(M^2+P^2)}-e^{-s(M^2+p_3^2+{\bf p}^2\frac{\tanh eBs}{eBs})}\right] ,\nonumber\\
&&F_{\rm m}=4\int\frac{d^3P}{(2\pi)^3}\int_0^\infty dse^{-s(M^2+p_3^2+{\bf p}^2\frac{\tanh eBs}{eBs})}+\Pi_{\rm m}(K).
\end{eqnarray}
Then we find that only the integral in $F_1$ is divergent and can be removed by coupling constant renormalization. To be consistent with the regularization scheme used in evaluating the effective potential, we introduce the cutoff $\Lambda$ for momentum $P$. Then we obtain
\begin{eqnarray}
F_1=\frac{1}{G}-\frac{\Lambda}{2\pi^2}+\frac{M}{\pi}=\frac{M}{\pi}-\frac{M_0}{\pi}{\rm sgn}(G-G_c).
\end{eqnarray}
The term $F_2$ is finite. Completing the integral over $P$ we get
\begin{eqnarray}
F_2=\frac{1}{2\pi^{3/2}}\int_0^\infty\frac{ds}{s^{3/2}}e^{-sM^2}\left(1-\frac{eBs}{\tanh eBs}\right)=
-\frac{E_{\rm B}}{2\pi^{3/2}}f_{3/2}\left(\frac{M}{E_{\rm B}}\right).
\end{eqnarray}

The term $F_{\rm m}$ is also finite. Therefore we can safely shift the integration variables. Making use of the identity
\begin{eqnarray}
\int_0^\infty dte^{-t\left[M^2+(p_3+k_3)^2+({\bf p}+{\bf k})^2\frac{\tanh eBt}{eBt}\right]}
\left[M^2+(p_3+k_3)^2+\frac{({\bf p}+{\bf k})^2}{\cosh^2 eBt}\right]=1,
\end{eqnarray}
we can express $F_{\rm m}$ in a symmetric form,
\begin{eqnarray}
F_{\rm m}=\frac{4}{(2\pi)^3}\int_0^\infty ds\int_0^\infty dt e^{-(s+t)M^2-{\cal R}(s,t)}\int_{-\infty}^\infty dq_1 \int_{-\infty}^\infty dq_2 \int_{-\infty}^\infty dq_3e^{-{\cal A}(q_1,q_2,q_3)}{\cal H}_{\rm m}(q_1,q_2,q_3),
\end{eqnarray}
where
\begin{eqnarray}
{\cal A}(q_1,q_2,q_3)=\frac{f_s+f_t}{eB}(q_1^2+q_2^2)+(s+t)q_3^2
\end{eqnarray}
and
\begin{eqnarray}
{\cal H}_{\rm m}(q_1,q_2,q_3)&=&\left(\frac{f_s^2+f_t^2}{2}-f_s^2f_t^2\right)(q_1^2+q_2^2)-f_sf_tq_3^2
+\frac{f_s^2+f_t^2-2f_s^2f_t^2+2(1-f_s^2)(1-f_t^2)f_sf_t}{2(f_s+f_t)^2}(k_1^2+k_2^2)\nonumber\\
&+&\frac{s^2+t^2+2(1+f_sf_t)st}{2(s+t)^2}k_3^2+M^2-\alpha_{\rm m}M^2(1+f_sf_t).
\end{eqnarray}
Completing the integral over $q_1,q_2,q_3$ we get
\begin{eqnarray}
&&\int_{-\infty}^\infty dq_1 \int_{-\infty}^\infty dq_2 \int_{-\infty}^\infty dq_3e^{-{\cal A}(q_1,q_2,q_3)}{\cal H}_{\rm m}(q_1,q_2,q_3)\nonumber\\
&=&\frac{\pi^{3/2}}{\sqrt{s+t}}\frac{eB}{f_s+f_t}\Bigg[\frac{eB}{f_s+f_t}\left(\frac{f_s^2+f_t^2}{2}-f_s^2f_t^2\right)-\frac{f_sf_t}{2(s+t)}
+M^2-\alpha_{\rm m}M^2(1+f_sf_t)\nonumber\\
&+&\frac{f_s^2+f_t^2-2f_s^2f_t^2+2(1-f_s^2)(1-f_t^2)f_sf_t}{2(f_s+f_t)^2}{\bf k}^2
+\frac{s^2+t^2+2(1+f_sf_t)st}{2(s+t)^2}k_3^2\Bigg].
\end{eqnarray}
Next, we define two new variables $z=s+t$ and $u=(s-t)/(s+t)$ and obtain
\begin{eqnarray}
F_{\rm m}(K)=\frac{eB}{2\pi^{3/2}}\int_0^1 du\int_0^\infty dz\sqrt{z}\ e^{-zM^2-\frac{z}{4}(1-u^2)k_3^2-{\cal C}_0{\bf k}^2}
\left({\cal C}_{\rm m}+{\cal C}_1{\bf k}^2+{\cal C}_2k_3^2\right),
\end{eqnarray}
where
\begin{eqnarray}
&&{\cal C}_0=\frac{\cosh{(eBz)}-\cosh{(eBzu)}}{2eB\sinh{(eBz)}},\ \ \ \ \ {\cal C}_{\rm m}
=\frac{eB}{2}\frac{\cosh{(eBz)}\cosh{(eBzu)}-1}{\sinh^2{(eBz)}}
-\left(\frac{1}{2z}+M^2\right)eB{\cal C}_0+\frac{(1-\alpha_{\rm m})M^2}{\tanh(eBz)},\nonumber\\
&&{\cal C}_1=\frac{3\cosh{(eBzu)}}{4\sinh{(eBz)}}+\frac{\left(2eB{\cal C}_0\right)^2-1}{4\tanh{(eBz)}},\ \ \ \ \
{\cal C}_2=\frac{1}{2\tanh{(eBz)}}-\frac{1+u^2}{4}eB{\cal C}_0.
\end{eqnarray}

At leading order in $1/N$, we find $F_1+F_2=0$ from the gap equation. The propagators of the collective modes are given by
\begin{eqnarray}
{\cal D}_{\sigma}(K)=\frac{1}{N}\frac{1}{F_{\sigma}(K)},\ \ \ \ \ \ {\cal D}_{\pi}(K)=\frac{1}{N}\frac{1}{F_{\pi}(K)}.
\end{eqnarray}
For the pionic excitation ($\alpha_{\rm m}=1$), we obtain
\begin{eqnarray}
F_\pi(0)&=&\frac{E_{\rm B}}{4\pi^{3/2}}\int_0^1 du\int_0^\infty dx\sqrt{x}\ e^{-\eta^2x}\left[\frac{\cosh x\cosh ux-1}{\sinh^2x}
-\frac{\cosh x-\cosh ux}{\sinh x}\left(\frac{1}{2x}+\eta^2\right)\right]\nonumber\\
&=&\frac{E_{\rm B}}{8\pi^{3/2}}\int_0^\infty \frac{dx}{x^{3/2}} e^{-\eta^2x}\left(1+2\eta^2x+\frac{x-2\eta^2x^2}{\tanh x}-\frac{2x^2}{\sinh^2x}\right).
\end{eqnarray}
Completing the integral over $x$, we find that $F_\pi(0)\equiv0$ for arbitrary nonzero value of $\eta$. Hence the Goldstone's theorem holds for arbitrary magnetic field. On the other hand, for vanishing magnetic field, the propagators reduce to \cite{NJL2DReview}
\begin{eqnarray}
{\cal D}_\sigma(K)=\frac{2\pi}{N}\frac{k}{(k^2+4M^2)\arctan\frac{k}{2M}},\ \ \ \ \ \
{\cal D}_\pi(K)=\frac{2\pi}{N}\frac{1}{k\arctan\frac{k}{2M}}, \label{propagatorB0}
\end{eqnarray}
where $k^2=k_3^2+{\bf k}^2$. Therefore, our results are consistent with the known expressions at $B=0$ \cite{NJL2DReview}.

To study the properties of the collective modes, we convert $k_3$ back to $-ik_0$. The velocity of the Goldstone mode can be determined by making use of the small momentum expansion,
\begin{eqnarray}
\frac{1}{N}{\cal D}_\pi^{-1}(K)=\xi_1{\bf k}^2-\xi_2 k_0^2+\cdots.
\end{eqnarray}
The expansion coefficients can be evaluated as
\begin{eqnarray}
\xi_1&=&\frac{1}{2\pi^{3/2}E_{\rm B}}\int_0^1 du\int_0^\infty dx\sqrt{x}\ e^{-\eta^2x}\Bigg\{\frac{3\cosh ux}{4\sinh x}
+\frac{1}{4\tanh x}\left[\left(\frac{\cosh x-\cosh ux}{\sinh x}\right)^2-1\right]\nonumber\\
&&\ \ \ \ \ \ \ \ \ \ \ \ \ \ \ \ -\frac{\cosh x-\cosh ux}{4\sinh x}
\left[\frac{\cosh x\cosh ux-1}{\sinh^2x}-\frac{\cosh x-\cosh ux}{\sinh x}\left(\frac{1}{2x}+\eta^2\right)\right]\Bigg\},\nonumber\\
&=&\frac{1}{32\pi^{3/2}E_{\rm B}}\int_0^\infty \frac{dx}{x^{3/2}} e^{-\eta^2x}\left(12x+4\eta^2x^2-3\frac{2\eta^2x+2x\coth x+1}{\tanh x}+3x\frac{2\eta^2x+4x\coth x-1}{\sinh^2x}\right)\nonumber\\
\xi_2&=&\frac{1}{2\pi^{3/2}E_{\rm B}}\int_0^1 du\int_0^\infty dx\sqrt{x}\ e^{-\eta^2x}\Bigg\{\frac{1}{2\tanh x}
-\frac{1+u^2}{8}\frac{\cosh x-\cosh ux}{\sinh x}\nonumber\\
&&\ \ \ \ \ \ \ \ \ \ \ \ \ \ \ \ -\frac{1-u^2}{8}
\left[\frac{\cosh x\cosh ux-1}{\sinh^2x}-\frac{\cosh x-\cosh ux}{\sinh x}\left(\frac{1}{2x}+\eta^2\right)\right]\Bigg\},\nonumber\\
&=&\frac{1}{48\pi^{3/2}E_{\rm B}}\int_0^\infty \frac{dx}{x^{5/2}} e^{-\eta^2x}\left(9+6\eta^2x+x\frac{2\eta^2x^3+9x^2-6\eta^2x-3}{\tanh x}
+2x^2\frac{x^2-3}{\sinh^2x}\right)\nonumber\\
\end{eqnarray}
The integral over $x$ in $\xi_1$ can be completed to get $\xi_1=1/(4\pi M)$ \cite{NJL-B02} which indicates that the dynamics of the pion mode is not suppressed by the magnetic field. The Goldstone mode velocity is given by $\upsilon_\pi=\sqrt{\xi_1/\xi_2}$. In Fig. \ref{fig2}, we show the results of $\upsilon_\pi$ for both the
subcritical and supercritical cases. For $G>G_c$, we have $\eta\rightarrow\infty$ and hence $\upsilon_\pi\rightarrow1$ for $B\rightarrow0$. While for
$G<G_c$, we have $\eta\rightarrow0$ and hence $\upsilon_\pi\rightarrow0$ for $B\rightarrow0$. In the large magnetic field limit,
the velocity approaches a universal limit. This limit velocity can be determined by using the result $\eta\rightarrow0.4460$ for $B\rightarrow\infty$. We obtain
\begin{eqnarray}
\lim_{B\rightarrow\infty}\upsilon_\pi=0.5875.
\end{eqnarray}

Next we determine the mass and spectral property of the sigma meson. To this end, we consider the case of ${\bf k}=0$. At vanishing magnetic field, the sigma meson is a slightly bound state with mass $m_\sigma=2M$ coincident with the two-fermion threshold \cite{NJL2DReview}. At nonzero magnetic field, the inverse of the sigma meson propagator at ${\bf k}=0$ can be evaluated as
\begin{eqnarray}
F_\sigma(k_0=\omega)&=&\frac{E_{\rm B}}{4\pi^{3/2}}\int_0^1 du\int_0^\infty dx\sqrt{x}\ e^{-\eta^2x[1-(1-u^2)b^2]}\Bigg[\frac{\cosh x\cosh ux-1}{\sinh^2x}+\frac{4\eta^2(1-b^2)}{\tanh x}\nonumber\\
&&\ \ \ \ \ \ \ \ \ \ +\left(\eta^2b^2(1+u^2)-\frac{1}{2x}-\eta^2\right)\frac{\cosh x-\cosh ux}{\sinh x}\Bigg],
\end{eqnarray}
where $b\equiv \omega/(2M)$. We note that the branching cut remains $\omega\in(2M,+\infty)$ and the two-fermion threshold is still $\omega_{\rm th}=2M$ at nonzero magnetic field. Therefore, the sigma meson is an unstable resonance if $m_\sigma>2M$ and a bound state if $m_\sigma<2M$. Actually, we can show that at $B\neq0$, the sigma meson is still a slightly bound state and its mass always coincides with the two-fermion threshold, i.e. $m_\sigma=2M$ for arbitrary magnetic field. The integral form of (45) is singular at $\omega=2M$ and its principal value is hard to obtain. We therefore turn to another form of $F_\sigma(\omega)$. By using the Ritus method which will be introduced in the next section, we can express $F_{\sigma}(\omega)$ as a summation over all Landau levels. The result is
\begin{eqnarray}
F_\sigma(\omega)&=&\frac{eB}{2\pi}\sum_{n=0}^{\infty}\alpha_n
\left[\frac{1}{\varepsilon_n}-\frac{8neB(1-\Theta(\omega-2\varepsilon_n))}{(4\varepsilon_n^2-\omega^2)\varepsilon_n}
+\frac{2(\omega^2-4M^2)\Theta(\omega-2\varepsilon_n)}{\omega(4\varepsilon_n^2-\omega^2)}\right], \label{ritussigma}
\end{eqnarray}
where $\varepsilon_n=\sqrt{M^2+2neB}$ and $\Theta(\omega-2\varepsilon_n)$ is the step function which equals $1$ for $\omega\ge2\varepsilon_n$ and equals $0$ for $\omega<2\varepsilon_n$. The degeneracy $\alpha_n=1$ for $n=0$ and $\alpha_n=2$ for $n\geq 1$. From this expression, we see obviously that $\omega=2M$ is always a pole of the sigma meson propagator. Therefore, the sigma meson is always a lightly bound state for arbitrary magnetic field, with its mass coincident with the two-fermion threshold.
\end{widetext}

In this section, we have studied the magnetic catalysis of dynamical chiral symmetry and its influence on the collective modes. While the magnetic catalysis \cite{NJL-B01,NJL-B02} and the properties of the collective modes \cite{NJL-B02} were studied long ago, here we have proposed a self-consistent scheme to study the properties of the collective modes. The propagators of the sigma and pion modes clearly recover the known results
at vanishing magnetic field [Eq. (\ref{propagatorB0})]. The mass of the sigma mode was investigated by using the method of low-energy expansion \cite{NJL-B02}. However, for heavy modes, the low-energy expansion becomes improper. Here, by using the explicit form of the inverse sigma propagator [Eq. (\ref{ritussigma})], we have shown that the sigma mode is a lightly bound state for arbitrary magnetic field, with its mass coincident with the two-fermion threshold.

Finally, we point out that the above scheme of evaluating the propagators of the collective modes has its advantage if we compute the next-to-leading
order corrections in $1/N$. The next-to-leading order contributions to the effective potential can be written as \cite{NJL-NLO}
\begin{eqnarray}
\Omega^{(1)}(M)=U_\sigma(M)+U_\pi(M),
\end{eqnarray}
where the two contributions $U_\sigma(M)$ and $U_\pi(M)$ read
\begin{eqnarray}
U_\sigma(M)&=&\frac{1}{2}\int\frac{d^3K}{(2\pi)^3}\ln\left[1+G\Pi_\sigma(K)\right],\nonumber\\
U_\pi(M)&=&\frac{1}{2}\int\frac{d^3K}{(2\pi)^3}\ln\left[1+G\Pi_\pi(K)\right].
\end{eqnarray}
To renormalize the total effective potential, it is natural to use the same cutoff $\Lambda$ to regularize the integrals over the Euclidean momenta $K$. Meanwhile, it is also convenient to separate $\Omega^{(1)}$ into a vacuum part and a $B$-dependent part. The next-to-leading corrections in $1/N$ enable us to quantitatively study the competition between the magnetic catalysis and the magnetic inhibition \cite{BInhibition} in the planar NJL model. The results will be reported elsewhere.

\section {Finite Temperature: Kosterlitz-Thouless transition} \label{s3}
From the properties of the collective modes at zero temperature, we find that the dynamics of the collective modes
is still (2+1)-dimensional even in the strong magnetic field limit. In the large-$N$ limit, phase fluctuations of the order
parameter are completely suppressed and the system undergoes a second-order phase transition at which the dynamical fermion mass
is generated. However, for finite $N$, the CMWH theorem forbids any long-range order and hence spontaneous breaking of
the U$(1)$ chiral symmetry at any nonzero temperature \cite{CMWH}. Since the system is still effectively (2+1)-dimensional, we expect that
there exists a phase transition of the Kosterlitz-Thouless (KT) type \cite{KT}. The KT transition temperature of the NJL$_3$ model at vanishing
magnetic field has been studied by Babaev \cite{Babaev}. In this section, we study the magnetic field dependence of the KT transition temperature.
Since we employ four-component spinor, we can introduce a chemical potential $\mu_5$ which corresponds to a chiral imbalance. To this end,
we add a chemical potential term $\mu_5\bar\psi\gamma^0\gamma^5\psi$. The meaning of the chiral imbalance or chiral chemical potential
becomes explicit if we define the left- and right-handed fermion fields as $\psi_{\rm L,R}=\frac{1}{2}(1\mp\gamma_5)\psi$. Then the
chiral chemical potential term becomes
\begin{eqnarray}
\mu_5\bar\psi\gamma^0\gamma^5\psi=\mu_5(\psi_{\rm R}^\dagger\psi_{\rm R}^{\phantom{\dag}}-\psi_{\rm L}^\dagger\psi_{\rm L}^{\phantom{\dag}}).
\end{eqnarray}
Therefore, $\mu_5$ is the chemical potential associated with the imbalance between the left- and right-handed fermions.

In some planar condensed matter systems such as graphene, $\mu_5$ corresponds to the chemical potential of doped Dirac electrons \cite{Electron2D}. To understand this, we introduce a new field $\Psi=\frac{1}{2}(1+\gamma_5)\psi+\frac{1}{2}(1-\gamma_5)\psi_c$ \cite{Babaev02}, where $\psi_c=C\bar{\psi}^{\rm T}$ with $C=-i\gamma^2\gamma^0$ being the charge conjugate matrix. The chemical potential term $\mu_5\bar\psi\gamma^0\gamma^5\psi$ turns to be the usual one $\mu_5\bar{\Psi}\gamma^0\Psi$. Meanwhile, we can show that the planar NJL model Eq. (1) is equivalent to the following BCS model of ultra-relativistic fermions \cite{He},
\begin{eqnarray}
{\cal L}_{\rm BCS}=\bar{\Psi}i\feyn{\partial}\Psi
+\frac{G}{2N}(\bar{\Psi}i\gamma_5\Psi_c)(\bar{\Psi}_ci\gamma_5\Psi).
\end{eqnarray}
Therefore, our studies in this section will also be relevant to the superconducting phenomenon of Dirac electrons in planar condensed matter systems.

\subsection{Phase Fluctuations and Kosterlitz-Thouless Transition}
At finite temperature, the partition function of the NJL$_3$ model is given by
\begin{eqnarray}
&&Z=\int[d\bar{\psi}][d\psi][d\sigma][d\pi]\exp\left\{\int_0^\beta d\tau\int d^2{\bf r} {\cal L}_{\rm eff}\right\},\nonumber\\
&&{\cal L}_{\rm eff}=\bar{\psi}(i\feyn{D}+\mu_5\gamma^0\gamma^5-\sigma-i\gamma_5\pi)\psi-\frac{N}{2G}(\sigma^2+\pi^2),\;\;
\end{eqnarray}
where $\beta=1/T$ with $T$ being the temperature. The KT transition temperature $T_{\rm KT}$ of the system can be
determined by studying the low-energy effective theory of the phase $\theta(x)$ of the order parameter field $\Delta(x)$,
which by employing the ``modulus-phase" variables \cite{Witten} is defined as
\begin{eqnarray}
\Delta(x)=\sigma(x)+i\pi(x)=\rho(x)e^{i\theta(x)}.
\end{eqnarray}
The order parameter field $\Delta(x)$ corresponds to the expectation value of the bilinear field $\bar{\Psi}_ci\gamma_5\Psi$ in the BCS Lagrangian Eq. (50). In terms of $\Delta(x)$, chiral symmetry can be written as $\Delta\rightarrow\Delta e^{2i\phi}$ or $\theta\rightarrow\theta+c$. In terms of the modulus-phase variables, the effective action reads
\begin{eqnarray}
\Gamma_{\rm eff}[\rho,\theta]&=&-N{\rm Trln}\left[i\feyn{D}+\mu_5\gamma^0\gamma^5-\rho(x)e^{i\gamma_5\theta(x)}\right]\nonumber\\
&&+N\int_0^\beta d\tau\int d^2{\bf r}\frac{\rho^2(x)}{2G}.
\end{eqnarray}

To study the KT transition, we need only to analyze the infrared behavior of the theory. To this end, we can just replace $\rho(x)$ by
its expectation value $\langle\rho(x)\rangle=M$ and neglect its fluctuations. Because of strong phase fluctuations, the expectation value of the order parameter always vanishes at finite temperature, i.e.,
\begin{eqnarray}
\langle\Delta(x)\rangle=\langle\rho(x)e^{i\theta(x)}\rangle=0.
\end{eqnarray}
Therefore, a nonzero expectation value $M$ does not break the chiral symmetry, in contrast to the zero temperature case. The effective potential for $M$ can be evaluated by setting $\theta=0$. We obtain
\begin{eqnarray}
\Omega(M)=\frac{N}{2G}M^2-\frac{N}{V_{2+1}}{\rm Trln}\left(i\feyn{D}+\mu_5\gamma^0\gamma^5-M\right).
\end{eqnarray}
Minimizing the effective potential, we obtain the expectation value $M$.

The infrared behavior of the theory is determined by the quasi-massless field $\theta$. The next step is to obtain an effective Hamiltonian for the phase field $\theta$. It is obvious that only the term proportional to $(\nabla\theta)^2$ is important, since other terms which have higher dimensions are suppressed in the infrared limit. We also note that terms like $\theta^4$ and $\theta^2(\nabla\theta)^2$ are forbidden by the chiral symmetry $\theta\rightarrow\theta+c$. Finally, the low-energy effective Hamiltonian of the theory can be expressed as
\begin{eqnarray}
H_{\rm eff}=\frac{J}{2}\int d^2{\bf r}\left[\nabla\theta(\bf r)\right]^2,
\end{eqnarray}
where $J$ is the stiffness of the phase fluctuations. This is nothing but the continuum version of the 2D XY model which was first used
to study the KT transition. The difference is that the phase stiffness $J$ here is not a constant but depends on temperature and other
parameters of the system, i.e.,
\begin{eqnarray}
J=J(T,M,\mu_5,E_{\rm B})
\end{eqnarray}
The critical temperature of the KT transition is then given by
\begin{eqnarray}
T_{\rm KT}=\frac{\pi}{2}J(T_{\rm KT},M,\mu_5,E_{\rm B}).
\end{eqnarray}
This equation should be solved together with the gap equation for $M$ to obtain the KT transition temperature $T_{\rm KT}$ at given external
parameters $E_{\rm B}$ and $\mu_5$.

At finite $N$, we will have three phases at nonzero temperature: (i) $0<T<T_{\rm KT}$--the low temperature quasi-ordered KT phase. In this phase,
the correlation function of the order parameter field decays algebraically at large distance ($|{\bf r}_1-{\bf r}_2|\rightarrow\infty$),
\begin{eqnarray}
\langle\Delta({\bf r}_1)\Delta({\bf r}_2)\rangle\sim|{\bf r}_1-{\bf r}_2|^{-\xi(T)}.
\end{eqnarray}
The correlation length $\xi$ in this phase can be shown to be $\xi(T)=T/(2\pi J)$. We therefore have quasi long-range order in this phase. It is well known that bound vortex-antivortex pairs will form in this phase. (ii)
$T_{\rm KT}<T<T^*$--the intermediate temperature pseudogap phase. In this phase,
the correlator decays exponentially,
\begin{eqnarray}
\langle\Delta({\bf r}_1)\Delta({\bf r}_2)\rangle\sim e^{-|{\bf r}_1-{\bf r}_2|/\xi(T)}.
\end{eqnarray}
In this phase, we have a nonzero modulus of the order parameter which plays the role of a local fermion mass. However, free vortices form and forbid chiral symmetry breaking. (iii) $T>T^*$--high temperature normal phase with vanishing modulus of the order parameter.

\subsection{The gap equation and phase stiffness }
There are two approaches to deal with the problem of a relativistic fermionic system in an external magnetic field. One is the famous Schwinger approach \cite{Schwinger} which puts the fermion propagator in the form of the integration of the auxiliary proper-time over a complex function, the other is Ritus method \cite{Ritus} which solves Dirac equation directly and finds the eigenfunctions and eigenvalues. For $\mu_5\neq0$ the generalization of the fermion propagator is obscure in Schwinger approach. We therefore employ the Ritus method to evaluate the gap equation and the phase stiffness $J$. There is a good example \cite{Warringa2012} showing how the Dirac equation with an constant external magnetic field can be solved by using the Ritus method in $3+1$ dimensions and the generalization to $2+1$ dimensions is straightforward.

In a uniform external magnetic field $B$, the Dirac equation in the mean-field approximation takes the form
\begin{eqnarray}\label{Dirac}
\left[i \gamma^0\partial_0-i\gamma^1\partial_1-i \gamma^2(\partial_2+i eB x_1)+\mu_5\gamma^0\gamma^5-M\right]\psi(x)=0.\;\;
\end{eqnarray}
Since the time dimension $x_0$ and the space dimension $x_2$ do not couple with the external magnetic field, the eigenfunctions should be proportional to the plane waves $e^{\pm \varepsilon^{\rm s}x_0+ip_2x_2}$. Therefore, the eigen solutions of the Dirac equation take the form
\begin{eqnarray}
\psi_{\rm s}^+(x)&=&e^{-i\varepsilon^{\rm s}x_0+ip_2x_2}G\left(x_1-\frac{p_2}{eB}\right)u_{\rm s},\nonumber\\
\psi_{\rm s}^-(x)&=&e^{i\varepsilon^{-\rm s}x_0+ip_2x_2}G\left(x_1-\frac{p_2}{eB}\right)\upsilon_{\rm s},
\end{eqnarray}
where $x=(x_0,x_1,x_2)$ and $s=\pm$ which are related to the chirality. Here $u_{\rm s}$ and $\upsilon_{\rm s}$ are spinors for particle and anti-particle solutions respectively and we take their momenta to be both $p_2$ for convenience. The $4\times4$ matrix $G(x_1-{p_2\over eB})$ is related to the Landau levels. Substituting these formal solutions into the Dirac equation, we obtain
\begin{eqnarray}
\left(\varepsilon^{\rm s}\gamma^0-i\gamma^1\partial_y+\gamma^2eBy+\mu_5\gamma^0\gamma^5-M\right)G(y)u_{\rm s}&=&0,\nonumber\\
\left(-\varepsilon^{-\rm s}\gamma^0-i\gamma^1\partial_y+\gamma^2eBy+\mu_5\gamma^0\gamma^5-M\right)G(y)\upsilon_{\rm s}&=&0,
\end{eqnarray}
where $y\equiv x_1-p_2/(eB)$. To get the solutions of $u_s$ and $\upsilon_s$ we use the Ritus Ansatz for the matrix $G(y)$,
\begin{eqnarray}\label{Gx}
(i\gamma^1\partial_y-\gamma^2eB y)G(y)=\lambda G(y)\gamma^2.
\end{eqnarray}
Without loss of generality we can choose the matrix $G(y)$ to be diagonal and commute with other terms. Then we get the equations for $u_s$ and $\upsilon_s$,
\begin{eqnarray}
\left(\varepsilon^{\rm s}\gamma^0-\lambda\gamma^2+\mu_5\gamma^0\gamma^5-M\right)u_{\rm s}&=&0,\nonumber\\
\left(-\varepsilon^{-\rm s}\gamma^0-\lambda\gamma^2+\mu_5\gamma^0\gamma^5-M\right)\upsilon_{\rm s}&=&0.
\end{eqnarray}

Let $G(y)={\rm diag}\left(g_1(y),g_2(y),g_3(y),g_4(y)\right)$, we get $g_2(y)=g_4(y)$ and $g_3(y)=g_1(y)$. The functions $g_1(y)$ and $g_4(y)$ are determined by the coupled equations
\begin{eqnarray}
\left(\partial_y+eBy\right)g_1(y)&=&\lambda g_4(y),\nonumber\\
\left(\partial_y-eBy\right)g_4(y)&=&-\lambda g_1(y).
\end{eqnarray}
Substituting one equation into the other, we obtain two decoupled equations,
\begin{eqnarray}
\left[-\partial_y^2+(eBy)^2\right]g_1(y)&=&(\lambda^2+eB)g_1(y),\nonumber\\
\left[-\partial_y^2+(eBy)^2\right]g_4(y)&=&(\lambda^2-eB)g_4(y).
\end{eqnarray}
Then we can write $\lambda^2=\lambda_n^2\equiv2neB$ with $n=0,1,2,\cdots$. The full solution of $G(y)$ can be found by using the fact that $g_1(y)$ and $g_4(y)$ must have the same value of $\lambda^2$. We obtain $g_1(y)=c_1\phi_n(y)$ and $g_4(y)=c_4\phi_{n-1}(y)$. The function $\phi_n(y)$ is given by
\begin{equation}
\phi_{n}(y)=\frac{1}{\sqrt{2^n n!}}\left(\frac{eB}{\pi}\right)^{1/4} H_n( \sqrt{eB}y)\exp \left(-\frac{1}{2}eBy^2\right),
\end{equation}
where $H_n(z)$ denotes a Hermite polynomial of degree $n$. For convenience we define $\phi_{-1}(y) = 0$. From Eq. (62) we get $c_1=c_4=1$.
The diagonal matrix $G(y)\equiv G_n(y)$ can be written in a compact form
\begin{eqnarray}
G_n(y)=\frac{1+i\gamma^1\gamma^2}{2}\phi_n(y)+\frac{1-i\gamma^1\gamma^2}{2}\phi_{n-1}(y).
\end{eqnarray}
According to the properties of the function $\phi_n(y)$, we have
\begin{eqnarray}
\sum_nG_n(y)G_n(y^\prime)&=&\delta(y-y^\prime),\nonumber\\
\int dy G_n(y)G_{m}(y)&=&\delta_{nm}.
\end{eqnarray}
The solutions of the eigen energy $\varepsilon^{\rm s}\equiv\varepsilon_n^{\rm s}$ are given by
\begin{eqnarray}
\varepsilon_n^{\rm s}=\sqrt{(\lambda_n+{\rm s}\mu_5)^2+M^2}.
\end{eqnarray}
For $n=0$, we have $\varepsilon_0^\pm\equiv\varepsilon_0=\sqrt{\mu_5^2+M^2}$. At vanishing $\mu_5$, it becomes $\varepsilon_n^\pm=\varepsilon_n=\sqrt{2neB+M^2}$. The solutions of the spinors $u_{\rm s}$ and $\upsilon_{\rm s}$ can be expressed as
\begin{widetext}
\begin{eqnarray}
u_{\rm s}={1\over\sqrt{2}}\left(\begin{array}{cccc} i\sqrt{\varepsilon_n^{\rm s}+\mu_5+{\rm s}\lambda_n} \\
{\rm s}\sqrt{\varepsilon_n^{\rm s}+\mu_5+{\rm s}\lambda_n} \\
\frac{iM}{\sqrt{\varepsilon_n^{\rm s}+\mu_5+{\rm s}\lambda_n}} \\ \frac{{\rm s}M}{\sqrt{\varepsilon_n^{\rm s}+\mu_5+{\rm s}\lambda_n}} \end{array}\right),\ \ \ \ \ \ \
\upsilon_{\rm s}={1\over\sqrt{2}}\left(\begin{array}{cccc} i\sqrt{\varepsilon_n^{-\rm s}-\mu_5+{\rm s}\lambda_n} \\
-{\rm s}\sqrt{\varepsilon_n^{-\rm s}-\mu_5+{\rm s}\lambda_n} \\
\frac{-iM}{\sqrt{\varepsilon_n^{-\rm s}-\mu_5+{\rm s}\lambda_n}} \\ \frac{{\rm s}M}{\sqrt{\varepsilon_n^{-\rm s}-\mu_5+{\rm s}\lambda_n}} \end{array}\right).
\end{eqnarray}
Using these results, we obtain
\begin{eqnarray}
u_{\rm s}u_{\rm s}^\dagger&=&{1\over2}\Big[\varepsilon_n^{\rm s}(1
-i{\rm s}\gamma^1\gamma^3)-(\mu_5+{\rm s}\lambda_n)(\gamma^5+{\rm s}\gamma^2\gamma^0)+M(\gamma^0+{\rm s}\gamma^5\gamma^2)\Big],\nonumber\\
\upsilon_{\rm s}\upsilon_{\rm s}^\dagger&=&{1\over2}\Big[\varepsilon_n^{-\rm s}(1
-i{\rm s}\gamma^1\gamma^3)+(\mu_5+{\rm s}\lambda_n)(\gamma^5+{\rm s}\gamma^2\gamma^0)-M(\gamma^0+{\rm s}\gamma^5\gamma^2)\Big].
\end{eqnarray}
These results are useful in evaluating the fermion and pion propagators.

To evaluate the gap equation for $M$, we need to evaluate the fermion Green's function. First, the retarded Green's function for $x_0-x'_0>0$ can be evaluated as
\begin{eqnarray}
{\cal S}_{\rm R}(x,x')&\equiv&\langle0|\psi(x)\bar{\psi}(x')|0\rangle=\sum_{{\rm s}=\pm}\sum_{n=0}^\infty\int_{-\infty}^{\infty}
\frac{dp_2}{2\pi}{1\over{2\varepsilon_n^{\rm s}}}[\psi_{\rm s}^+(x)][\psi_{\rm s}^+(x')]^\dagger\gamma^0\nonumber\\
&=&\sum_{{\rm s}=\pm}\sum_{n=0}^\infty\int_{-\infty}^{\infty}
\frac{dp_2}{2\pi}{1\over{2\varepsilon_n^{\rm s}}}e^{-i\varepsilon_n^{\rm s}(x_0-x'_0)+ip_2(x_2-x'_2)}
G_n\left(x_1-{p_2\over eB}\right)u_{\rm s}u_{\rm s}^\dagger G_n\left(x_1^\prime-{p_2\over eB}\right)\gamma^0\nonumber\\
&=&\sum_{{\rm s}=\pm}\sum_{n=0}^\infty\int_{-\infty}^{\infty}
\frac{dp_2}{2\pi}{1\over{4\varepsilon_n^{\rm s}}}e^{-i\varepsilon_n^{\rm s}(x_0-x'_0)+ip_2(x_2-x'_2)}
\Bigg\{G_n\left(x_1-{p_2\over eB}\right)G_n\left(x_1^\prime-{p_2\over eB}\right)\left[\varepsilon_n^{\rm s}\gamma^0+
(\mu_5+{\rm s}\lambda_n)\gamma^0\gamma^5+M\right]\nonumber\\
&&+G_n\left(x_1-{p_2\over eB}\right)\gamma^1G_n\left(x_1^\prime-{p_2\over eB}\right)\left({\gamma^1}\right)^\dagger
\left[\varepsilon_n^{\rm s}\gamma^0-(\mu_5+{\rm s}\lambda_n)\gamma^0\gamma^5-M\right](-is\gamma^1\gamma^3)\Bigg\}.
\end{eqnarray}
Therefore, the Feynman Green's function is given by
\begin{eqnarray}
{\cal S}_{\rm F}(x,x')&=&\sum_{{\rm s}=\pm}\sum_{n=0}^\infty\int_{-\infty}^{\infty}\frac{dp_0}{2\pi}\int_{-\infty}^{\infty}
\frac{dp_2}{2\pi}{i\over2[p_0^2-(\varepsilon_n^{\rm s})^2]}e^{-ip_0(x_0-x'_0)+ip_2(x_2-x'_2)}\nonumber\\
&&\times\Bigg\{G_n\left(x_1-{p_2\over eB}\right)G_n\left(x_1^\prime-{p_2\over eB}\right)
\left[p_0\gamma^0+(\mu_5+{\rm s}\lambda_n)\gamma^0\gamma^5+M\right]\nonumber\\
&&+G_n\left(x_1-{p_2\over eB}\right)\gamma^1G_n\left(x_1^\prime-{p_2\over eB}\right)\left({\gamma^1}\right)^\dagger
\left[p_0\gamma^0-(\mu_5+{\rm s}\lambda_n)\gamma^0\gamma^5-M\right](-is\gamma^1\gamma^3)\Bigg\}.
\end{eqnarray}
At finite temperature, we replace $p_0\rightarrow i\omega_\nu\equiv i(2\nu+1)\pi T$ and $\int_{-\infty}^{\infty}\frac{dp_0}{2\pi}\rightarrow T\sum_{\nu=-\infty}^\infty$. Then the gap equation at finite temperature is given by the self-consistent Green's function relation
\begin{eqnarray}\label{Gap}
{N\over G}M={\rm Tr}{\cal S}_{\rm F}(x,x)=N\sum_{{\rm s}=\pm}\sum_{n=0}^\infty T\sum_{\nu=-\infty}^{\infty}\int_{-\infty}^{\infty}\frac{dp_2}{2\pi}
{ M\over{2[(i\omega_\nu)^2-(\varepsilon_n^{\rm s})^2]}}{\rm Tr}\left[G_n\left(x_1-{p_2\over eB}\right)G_n\left(x_1-{p_2\over eB}\right)\right].
\end{eqnarray}
Obviously, the gap equation is essentially the extreme condition $\partial\Omega/\partial M=0$. Completing the integral over $p_2$ and the summation over the Matsubara frequency $i\omega_\nu$ and employing the same regularization method as in Sec. II, we obtain
\begin{eqnarray}\label{Gap}
\frac{1}{N}\frac{\partial\Omega}{\partial M}=\frac{M^2}{\pi}-\frac{MM_0}{\pi}{\rm sgn}(G-G_c)-\frac{eB}{2\pi}\frac{\eta f_{3/2}\left(\eta\right)}{\sqrt{\pi}}
-\frac{eB}{4\pi}\sum_{{\rm s}=\pm}\sum_{n=0}^\infty\alpha_n M\left[\frac{1-2n_{\rm F}(\varepsilon_n^{\rm s})}{\varepsilon_n^{\rm s}}-\frac{1}{\varepsilon_n}\right]=0,
\end{eqnarray}
where $n_{\rm F}(x)=1/(1+e^{\beta x})$ is the Fermi distribution function. Note that unlike the zero temperature case, at finite temperature $M=0$ is always an extreme of the effective potential, i.e.,
\begin{eqnarray}\label{Gap}
\frac{1}{N}\frac{\partial\Omega}{\partial M}\bigg|_{M=0}=0.
\end{eqnarray}

To evaluate the phase stiffness, we need to evaluate the inverse of the pion propagator and make the small momentum expansion. We have
\begin{eqnarray}
J=M^2\lim_{{\bf k}\rightarrow 0}\frac{{\cal D}^{-1}_\pi(k_0=0,{\bf k})}{{\bf k}^2}.
\end{eqnarray}
The inverse of the pion propagator in coordinate representation can be evaluated as
\begin{eqnarray}
{\cal D}_{\pi}^{-1}(x,x')&=&{N\over G}\delta(x-x')-i{\rm Tr}[\gamma_5{\cal S}_{\rm F}(x,x')\gamma_5{\cal S}_{\rm F}(x',x)]\nonumber\\
&=&{N\over G}\delta(x-x')+\frac{i}{2}\sum_{{\rm s,t}=\pm}\sum_{n,m=0}^\infty\int_{-\infty}^\infty\frac{dp_0}{2\pi}\int_{-\infty}^\infty\frac{dp_0^\prime}{2\pi}\int_{-\infty}^\infty\frac{dp_2}{2\pi}
\int_{-\infty}^\infty\frac{dp_2^\prime}{2\pi}\frac{e^{-i(p_0-p_0^\prime)(x_0-x_0^\prime)+i(p_2-p_2^\prime)(x_2-x_2^\prime)}}
{[p_0^2-(\varepsilon_n^{\rm s})^2][p_0^{\prime 2}-(\varepsilon_m^{\rm t})^2]}\nonumber\\
&&\times \left[M^2-p_0p_0^\prime+(\mu_5+{\rm s}\lambda_n)(\mu_5+{\rm t}\lambda_m)\right]
\Bigg[\phi_n\left(x_1-{p_2\over eB}\right)\phi_n\left(x_1^\prime-{p_2\over eB}\right)\phi_m\left(x_1-{p_2^\prime\over eB}\right)\phi_m\left(x_1^\prime-{p_2^\prime\over eB}\right)\nonumber\\
&&+\phi_{n-1}\left(x_1-{p_2\over eB}\right)\phi_{n-1}\left(x_1^\prime-{p_2\over eB}\right)\phi_{m-1}\left(x_1-{p_2^\prime\over eB}\right)\phi_{m-1}\left(x_1^\prime-{p_2^\prime\over eB}\right)\nonumber\\
&&+{\rm st}\phi_n\left(x_1-{p_2\over eB}\right)\phi_{n-1}\left(x_1^\prime-{p_2\over eB}\right)\phi_m\left(x_1-{p_2^\prime\over eB}\right)\phi_{m-1}\left(x_1^\prime-{p_2^\prime\over eB}\right)\nonumber\\
&&+{\rm st}\phi_{n-1}\left(x_1-{p_2\over eB}\right)\phi_n\left(x_1^\prime-{p_2\over eB}\right)\phi_{m-1}\left(x_1-{p_2^\prime\over eB}\right)\phi_m\left(x_1^\prime-{p_2^\prime\over eB}\right)\Bigg].
\end{eqnarray}
The momentum representation of the inverse pion propagator can be obtained by Fourier transformation. We note that ${\cal D}_{\pi}^{-1}(x,x^\prime)$ should only be a function of $x-x^\prime$. Then we obtain
\begin{eqnarray}
{\cal D}_\pi^{-1}(k_0,{\bf k})=\int d^3x e^{i[k_0(x_0-x_0^\prime)-k_1(x_1-x_1^\prime)-k_2(x_2-x_2^\prime)]}D_{\pi}^{-1}(x-x^\prime).
\end{eqnarray}
After a lengthy calculation, the phase stiffness at finite temperature can be expressed as
 \begin{eqnarray}
\frac{J}{N}&=&{M^2\over8\pi}\sum_{{\rm s,t}=\pm}\sum_{n,m=0}^\infty T\sum_{\nu=-\infty}^\infty
\frac{(i\omega_\nu)^2-M^2-\left(\mu_5+{\rm s}\lambda_n\right)\left(\mu_5+{\rm t}\lambda_m\right)}
{\left[(i\omega_\nu)^2-(\varepsilon_{n}^{\rm s})^2\right]\left[(i\omega_\nu)^2-(\varepsilon_{m}^{\rm t})^2\right]}\nonumber\\
&&\times\left[(\sqrt{n}+{\rm st}\sqrt{n-1})^2\delta_{m,n-1}(\alpha_n-1)+(\sqrt{n}+{\rm st}\sqrt{n+1})^2\delta_{m,n+1}
-4n(1+{\rm st})\delta_{mn}(\alpha_n-1)-\delta_{mn}\delta_{n0}\right].
\end{eqnarray}
Completing the summation over ${\rm t}$ and $m$ and the summation over the Matsubara frequency , we finally obtain
\begin{eqnarray}
\frac{J}{N}&=&\frac{M^2}{4\pi}\frac{\mu_5^2-eB}{2\mu_5^2-eB}\frac{1}{\varepsilon_0}\tanh{\frac{\varepsilon_0}{2T}}
-\frac{M^2}{8\pi}\sum_{{\rm s}=\pm}\frac{\lambda_1+{\rm s}\mu_5}{\lambda_1+2{\rm s}\mu_5}\frac{1}{\varepsilon_1^{\rm s}}
\tanh{\frac{\varepsilon_1^{\rm s}}{2T}}\nonumber\\
&&+\frac{M^2}{16\pi}\sum_{{\rm s}=\pm}\sum_{n=1}^\infty\Bigg\{\frac{2n(eB)^2+eB\mu_5(2\mu_5+{\rm s}\lambda_n)-8n\mu_5^2(\mu_5+{\rm s}\lambda_n)^2}
{(eB)^2-4\mu_5^2(\mu_5+{\rm s}\lambda_n)^2}\frac{2}{\varepsilon_n^{\rm s}}\tanh{\varepsilon_n^{\rm s}\over 2T}\nonumber\\
&&-\frac{(\mu_5+{\rm s}\lambda_{n+1})[2(2n+1)\mu_5+{\rm s}\lambda_{n+1}]}
{2\mu_5^2+2{\rm s}\mu_5\lambda_{n+1}+eB}\frac{1}{\varepsilon_{n+1}^{\rm s}}\tanh{\varepsilon_{n+1}^{\rm s}\over 2T}
-\frac{(\mu_5+{\rm s}\lambda_{n-1})[2(2n-1)\mu_5-{\rm s}\lambda_{n-1}]}
{2\mu_5^2+2{\rm s}\mu_5\lambda_{n-1}-eB}\frac{1}{\varepsilon_{n-1}^{\rm s}}\tanh{\varepsilon_{n-1}^{\rm s}\over 2T}\Bigg\}.
\end{eqnarray}
 Using the same method, the sigma meson propagator ${\cal D}_\sigma(k_0,{\bf k})$ can be evaluated. The result for ${\bf k}=0$ and $T=0$ has been presented in Sec. II.
\end{widetext}

\subsection{Results for $\mu_5=0$}
At vanishing chiral imbalance, $\mu_5=0$, we have
\begin{eqnarray}
\frac{1}{N}\frac{\partial\Omega}{\partial M}&=&\frac{M^2}{\pi}-\frac{MM_0}{\pi}{\rm sgn}(G-G_c)-\frac{eB}{2\pi}\frac{\eta f_{3/2}\left(\eta\right)}{\sqrt{\pi}}\nonumber\\
&&+\frac{eB}{\pi}n_{\rm F}(M)+\frac{2eB}{\pi}\sum_{n=1}^\infty\frac{M}{\varepsilon_n}n_{\rm F}(\varepsilon_n).
\end{eqnarray}
Unlike the zero temperature case, at finite but low temperature $T<T^*$, the gap equation $\partial\Omega/\partial M=0$ has two solutions $M=0$ and $M\neq0$. The solution $M=0$ corresponds to a maximum. At $T=T^*$, the two extremes merge. The temperature $T^*$ is then determined by
\begin{eqnarray}
\frac{1}{N}\frac{\partial^2\Omega}{\partial M^2}\bigg|_{M=0}=0.
\end{eqnarray}
We obtain
\begin{eqnarray}
\gamma-\frac{1}{g_{\rm B}}{\rm sgn}(G-G_c)=\frac{1}{4t^*}-\sum_{n=1}^\infty\sqrt{\frac{2}{n}}\frac{1}{e^{\sqrt{2n}/t^*}+1},
\end{eqnarray}
where $t^*=T^*/\sqrt{eB}$ and
\begin{eqnarray}
\gamma=\lim_{\eta\rightarrow0}\frac{2\eta^2f_{1/2}(\eta)-f_{3/2}(\eta)}{2\sqrt{\pi}}\simeq1.0326.
\end{eqnarray}
In the strong magnetic field limit, we have
\begin{eqnarray}
\lim_{B\rightarrow\infty}\frac{T^*}{\sqrt{eB}}=0.2411.
\end{eqnarray}

The phase stiffness at $\mu_5=0$ can be simplified to
\begin{eqnarray}
J={N\over4\pi}M\tanh{M\over2T}.
\end{eqnarray}
The KT transition temperature $T_{\rm KT}$ is determined by the equation
\begin{eqnarray}
\frac{M}{T_{\rm KT}}\tanh{\frac{M}{2T_{\rm KT}}}=\frac{8}{N}
\end{eqnarray}
together with the gap equation
\begin{eqnarray}
\frac{1}{N}\frac{\partial\Omega}{\partial M}\bigg|_{T=T_{\rm KT}}=0.
\end{eqnarray}
For $N\rightarrow\infty$, we have $M(T=T_{\rm KT})\rightarrow0$ and therefore the KT transition temperature
coincides with $T^*$. For finite $N$ we obtain
\begin{eqnarray}
M(T=T_{\rm KT})=x_0(N)T_{\rm KT},
\end{eqnarray}
where $x_0(N)$ is the solution of the equation $x\tanh(x/2)=8/N$. Then the KT transition temperature is determined by
\begin{eqnarray}
&&x_0t_{\rm KT}-\frac{1}{g_{\rm B}}{\rm sgn}(G-G_c)-\frac{1}{2\sqrt{\pi}}f_{3/2}\left(x_0t_{\rm KT}\right)\nonumber\\
&=&-\sum_{n=0}\frac{\alpha_n}{\sqrt{x_0^2t_{\rm KT}^2+2n}}\frac{1}{e^{\sqrt{x_0^2t_{\rm KT}^2+2n}/t_{\rm KT}}+1},
\end{eqnarray}
where $t_{\rm KT}=T_{\rm KT}/\sqrt{eB}$. In the strong magnetic field limit, $t_{\rm KT}$ also approaches a constant depending
on the value of $N$. Therefore, with increasing magnetic field, the domain of the pseudogap phase $T_{\rm KT}<T<T^*$ is enlarged.
\begin{figure}[!htb]
\begin{center}
\includegraphics[width=8cm]{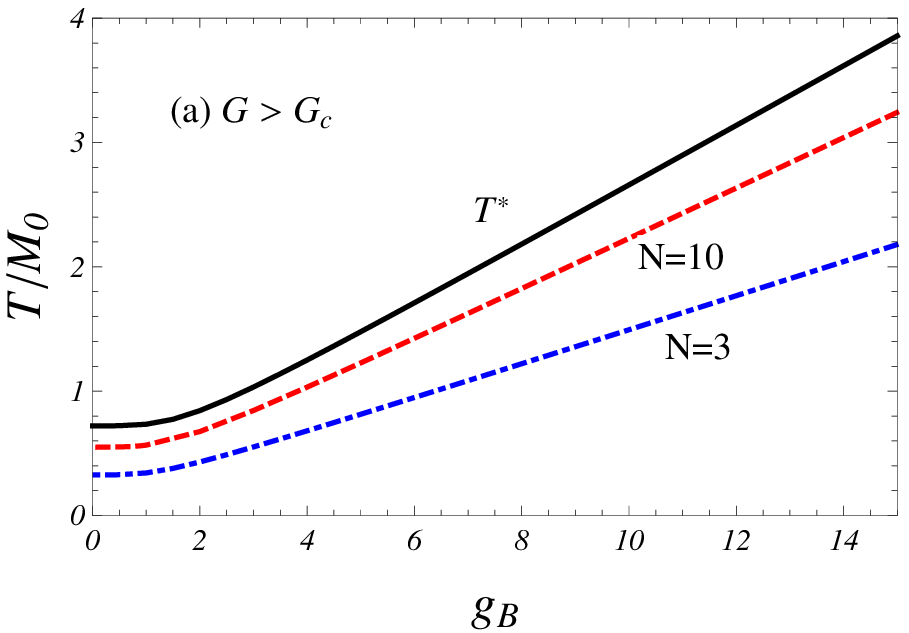}
\includegraphics[width=8cm]{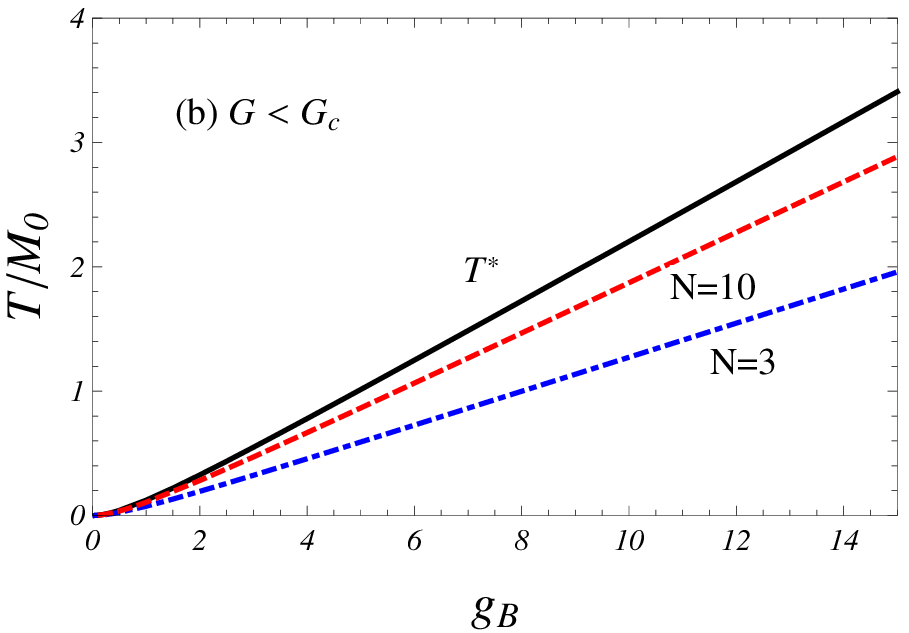}
\caption{(Color-online) The KT transition temperature $T_{\rm KT}$ and the mass melting temperature $T^*$ as a function of $g_B=E_{\rm B}/M_0$
for $G>G_c$ (a) and $G<G_c$ (b). For KT transition temperature, the results for $N=3$ and $N=10$ are shown with blue
dot-dashed lines and red dashed lines, respectively. \label{fig2}}
\end{center}
\end{figure}

\begin{figure}[!htb]
\begin{center}
\includegraphics[width=8cm]{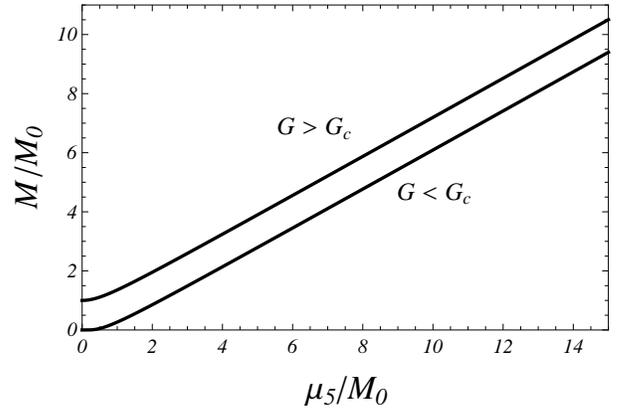}
\caption{The dynamical fermion mass $M$ (scaled by $M_0$) as a function of $\mu_5/M_0$ for $G>G_c$ and $G<G_c$. \label{fig3}}
\end{center}
\end{figure}

The numerical results for $T^*$ and the KT transition temperature $T_{\rm KT}$ for $N=3$ and $N=10$ are shown in Fig. 3. Note that
$T^*$ is independent of $N$. We find that for a given value of $N$, the pseudogap domain $T_{\rm KT}<T<T^*$ becomes larger and larger with
increasing magnetic field.

\subsection{Results for $\mu_5\neq0$}

For nonvanishing $\mu_5$, the effective potential at $B=0$ and $T=0$ can be evaluated as
\begin{eqnarray}
\frac{1}{N}\Omega(M)&=&\frac{1}{N}\Omega(M,\mu_5=0)+\int_0^\infty {p{\rm d}p\over2\pi}\Bigg[2\sqrt{p^2+M^2}\nonumber\\
&&-\sqrt{(p-\mu_5)^2+M^2}-\sqrt{(p+\mu_5)^2+M^2}\Bigg].\;
\end{eqnarray}
The fermionic excitation spectra $\sqrt{(p\pm\mu_5)^2+M^2}$ mean that the chiral imbalance $\mu_5$ actually plays the role of
a Fermi surface of left- or right-handed fermions. Completing the integral over $p$, we obtain
\begin{eqnarray}
\frac{1}{N}\Omega(M)&=&-\frac{M_0M^2}{2\pi}{\rm sgn}(G-G_c)+\frac{(2M^2-\mu_5^2)\sqrt{M^2+\mu_5^2}}{6\pi}\nonumber\\
&&-\frac{\mu_5M^2}{2\pi}\ln\frac{\sqrt{M^2+\mu_5^2}+\mu_5}{M}.
\end{eqnarray}
For $\mu_5\neq0$, we find that
\begin{eqnarray}
\frac{1}{N}\frac{\partial\Omega}{\partial M}\bigg|_{M=0}=0,\ \ \ \frac{1}{N}\frac{\partial^2\Omega}{\partial M^2}\bigg|_{M=0}=-\infty.
\end{eqnarray}
Therefore, the minimum of the effective potential is always located at $M\neq0$, no matter $G>G_c$ or $G<G_c$.
The numerical results are shown in Fig. 3. It is obvious that $\mu_5$ also catalyzes dynamical chiral symmetry breaking because of the Fermi surface effect. We note that the above result is similar to the chemical potential effect on the superconducting phenomenon of Dirac electrons in planar condensed matter systems \cite{Electron2D}.

Now we turn on the magnetic field. The temperature $T^*$ is determined by
\begin{eqnarray}
&&\gamma-\frac{1}{g_{\rm B}}{\rm sgn}(G-G_c)=\frac{1}{2\delta}\tanh\frac{\delta}{2t^*}+\sum_{n=1}^\infty\frac{\delta^2}{\sqrt{2n}(2n-\delta^2)}\nonumber\\
&&-2\sum_{n=1}^\infty\frac{\sqrt{2n}+\sqrt{2n}e^{\sqrt{2n}/t^*}\cosh\frac{\delta}{t^*}+\delta e^{\sqrt{2n}/t^*}\sinh\frac{\delta}{t^*}}
{(2n-\delta^2)(e^{\sqrt{8n}/t^*}+2e^{\sqrt{2n}/t^*}\cosh\frac{\delta}{t^*}+1)},\;\;
\end{eqnarray}
where $\delta=\mu_5/\sqrt{eB}$. Note that the singularities at $\delta=\sqrt{2n}$ are removable. The KT transition temperature $T_{\rm KT}$
is determined by solving the equation $T_{\rm KT}=\pi J/2$ together with the gap equation Eq. (74). We shall focus on the case $G>G_c$ and
$N=3$. The result for $G<G_c$ is similar because $\mu_5$ catalyzes dynamical chiral symmetry breaking.

The numerical results of $T^*$, $T_{\rm KT}$, and $M_{\rm KT}\equiv M(T=T_{\rm KT})$ for $\mu_5/M_0=5$ and $\mu_5/M_0=10$ are shown in Fig. 4. We find that there exists a regime of the magnetic field where these quantities first decrease and then increase, in contrast to the the case of $\mu_5=0$ where these quantities always increase with $\sqrt{eB}$. This phenomenon is more visible for the KT transition temperature $T_{\rm KT}$ and $M_{\rm KT}$. The decreasing behavior is therefore confusing since we have shown that either $B\neq0$ or $\mu_5\neq0$
enhances dynamical chiral symmetry breaking.

To understand the decreasing behavior or inverse magnetic catalysis of the transition temperatures, we note that the chiral imbalance $\mu_5$ plays the role of an effective Fermi surface. It is well-known that the combined effect of Fermi surface and magnetic field leads to the famous
de Haas--van Alphen (dHvA) oscillation \cite{dHvA}. The dHvA effect was first found in nonrelativistic systems such as metallic materials.
It was also found to exist in relativistic dense matter, such as dense quark matter \cite{NJL,dHvA,CSC-dHvA}. In the present system, we only find a minimum rather than multiple oscillations. As we will show in the following, this is because only the first excited Landau-Level is effective for the dHvA effect in the present system.

\begin{figure}[!htb]
\begin{center}
\includegraphics[width=8cm]{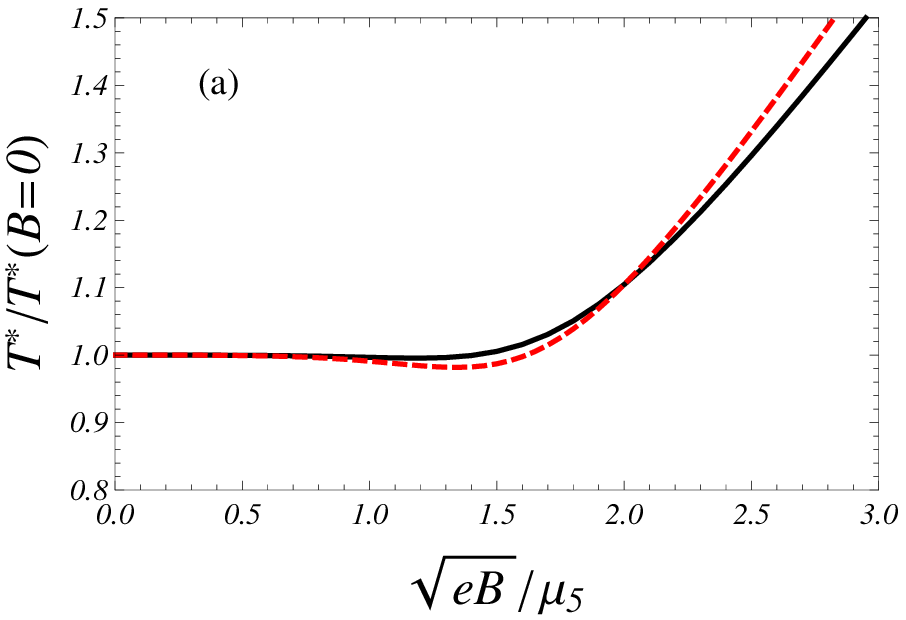}
\includegraphics[width=8cm]{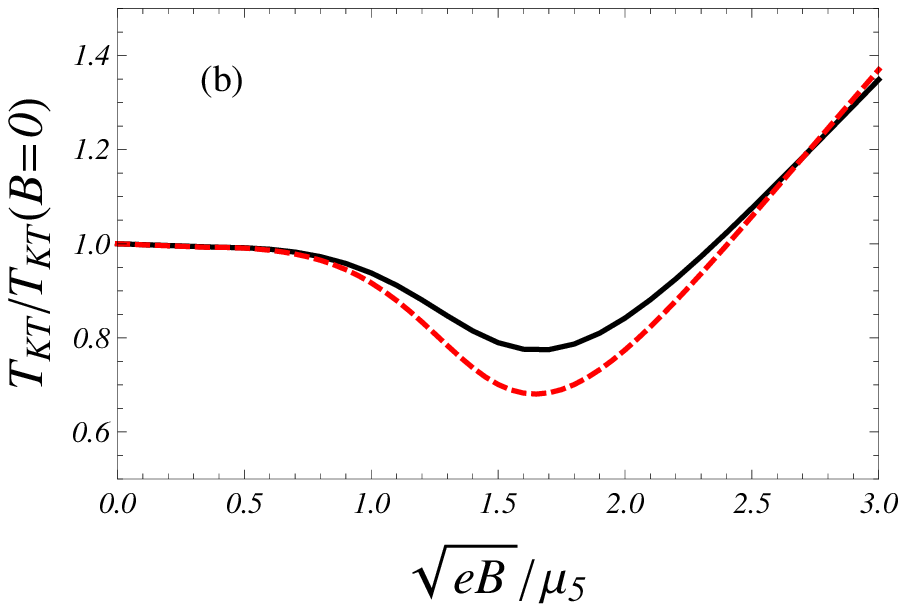}
\includegraphics[width=8cm]{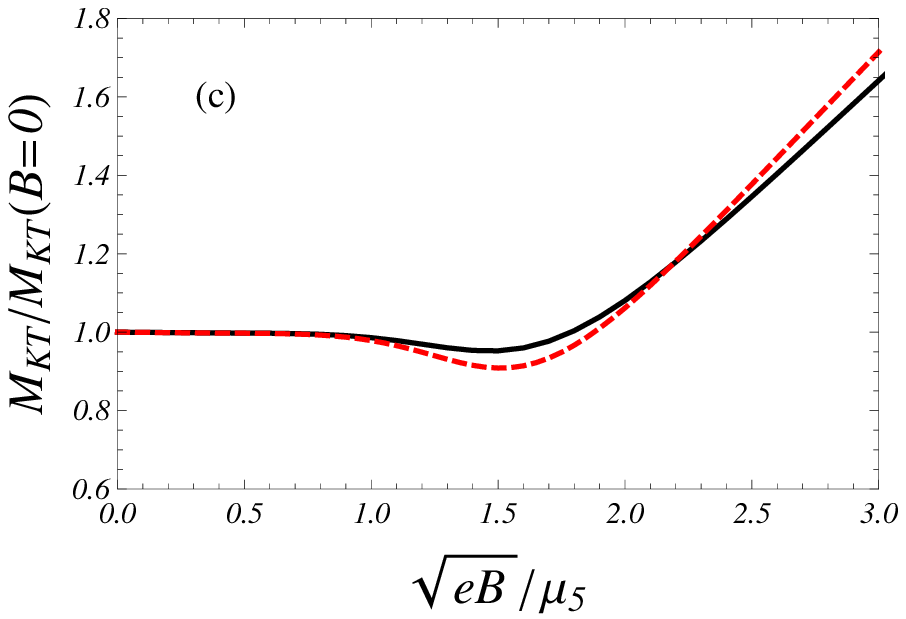}
\caption{(Color-online) Dependence of the mass melting temperature $T^*$ (a), KT transition temperature $T_{\rm KT}$ (b), and the effective mass $M_{\rm KT}$ at $T=T_{\rm KT}$ (c) on the magnetic field. $T^*$, $T_{\rm KT}$, and $M_{\rm KT}$ are all scaled by their values at $B=0$. The magnetic field strength is denoted by $\sqrt{eB}/\mu_5$. The black solid lines and the red dashed lines show the results for $\mu_5/M_0=5$ and $\mu_5/M_0=10$, respectively. \label{fig4}}
\end{center}
\end{figure}

To understand the dHvA effect quantitatively in the present system, we note that the dHvA effect is dominated by the terms which contain the negative branch of the excitation spectra, i.e.,
\begin{eqnarray}
\varepsilon_n^-=\sqrt{\left(\sqrt{2neB}-\mu_5\right)^2+M^2}
\end{eqnarray}
with $n\geq 1$. The dHvA oscillations are expected to occur when $\sqrt{eB}\sim \mu_5/\sqrt{2n}$. However, we find that the oscillations corresponding to $n\geq2$ are absent and only the one corresponding to $n=1$ exists in the present system. To understand this fact, we discuss three regimes of $\sqrt{eB}$.
\\ ({\bf A}) Weak magnetic field. This is roughly the regime $0<\sqrt{eB}/\mu_5<1/\sqrt{2}$. In this regime we expect that the Landau levels with $n\geq2$ will induce dHvA oscillations. However, $\mu_5$ dominates the behavior of this weak magnetic field regime. As a result, we obtain a plateau structure of $T_{\rm KT}$, $M_{\rm KT}$, and $T^*$ in this regime.
\\ ({\bf B}) Intermediate magnetic field. It corresponds roughly to the regime $1/\sqrt{2}<\sqrt{eB}/\mu_5<2$. We find that the first excited Landau-Level becomes effective and induces dHvA oscillation. The excitation spectrum of the first excited Landau-Level is $\varepsilon_1^-=\sqrt{(\sqrt{2eB}-\mu_5)^2+M^2}$. Because of the dHvA effect induced by the interplay between the magnetic field and the Fermi surface, the KT transition temperature (as well as $M_{\rm KT}$ and $T^*$) first decreases and then increases, inducing a minimum at the middle of this regime. The dHvA oscillation is more visible for the KT transition temperature, since it depends not only on the gap equation but also on the phase stiffness.
\\ ({\bf C}) Strong magnetic field. At large $\sqrt{eB}$, roughly corresponding to $\sqrt{eB}/\mu_5>2$, only the lowest Landau-Level is effective
and the magnetic catalysis effect dominates the behavior of the system. As a result, the KT transition temperature becomes nearly an increasing function of $\sqrt{eB}$. In the strong magnetic field limit we have $T_{\rm KT}\propto\sqrt{eB}$, because the influence of $\mu_5$ can be safely neglected.

In this section, we have studied the influence of a constant external magnetic field on the KT transition temperature. In the absence of chiral
imbalance $\mu_5$, we find that the KT transition temperature $T_{\rm KT}$ as well as the mass melting temperature $T^*$ is a monotonically increasing function of $\sqrt{eB}$. For a given value of $N$, the pseudogap region becomes larger for stronger magnetic field.
In the presence of chiral imbalance $\mu_5$, however, the KT transition temperature $T_{\rm KT}$ as well as the mass melting temperature $T^*$ goes non-monotonically with $\sqrt{eB}$. This behavior is similar to the inverse magnetic catalysis of the QCD
chiral transition temperature \cite{Lattice-B01,Lattice-B02}. In the present planar NJL model, it is evident that the non-monotonic behavior
of the KT transition temperature is actually a de Haas--van Alphen oscillation phenomenon induced by the interplay between the magnetic field and the chiral imbalance.

\section {Summary} \label{s4}
In the first part of this work we investigated the collective modes associated with the dynamical chiral symmetry breaking in a constant magnetic field in the (2+1)-dimensional Nambu--Jona-Lasinio model with continuous U(1) chiral symmetry. We introduced a self-consistent scheme to evaluate the propagators of the collective modes at the leading order in $1/N$. The scheme is proper to study the next-to-leading order corrections in $1/N$. We analytically proved that the sigma mode is always a lightly bound state with its mass coincident with the two-fermion threshold for arbitrary strength of the magnetic field. Because the dynamics of the collective modes is always 2+1 dimensional, the finite temperature transition should be of the KT type for finite $N$.

We also investigated the KT transition temperature $T_{\rm KT}$ as well as the mass melting temperature $T^*$ in a constant magnetic field and with an axial chemical potential $\mu_5$ in the second part of this work. The expression of the phase stiffness was derived by using the Ritus method. For vanishing chiral asymmetry $\mu_5$, we found that the pseudogap region $T_{\rm KT}<T<T^*$ is enlarged with increasing strength of the magnetic field. For nonzero $\mu_5$, we showed that it can lead to inverse magnetic catalysis of the KT transition temperature in 2+1 dimensions. This phenomenon can be attributed to the de Haas--van Alphen oscillation induced by the interplay between the magnetic field and Fermi surface. These results are also relevant to the superconducting phenomenon of Dirac electrons in planar condensed matter systems, such as graphene layers.

{\bf Acknowledgments:} We thank Dirk Rischke for helpful discussions and Igor Shovkovy for useful communications. Gaoqing Cao and Pengfei Zhuang are supported by the NSFC under grant No. 11335005 and the MOST under grant Nos. 2013CB922000 and 2014CB845400. Lianyi He is supported by the Department of Energy Nuclear Physics Office, by the topical collaborations on Neutrinos and Nucleosynthesis, and by Los Alamos National Laboratory. He also acknowledges the support from the Helmholtz International Center for FAIR within the framework of the LOEWE program launched by the State of Hesse in the early stage of this work.

\end{document}